
\magnification \magstep1
\raggedbottom
\openup 4\jot
\voffset6truemm
\headline={\ifnum\pageno=1\hfill\else
\hfill {\it The Impact of Quantum Cosmology on
Quantum Field Theory} \hfill \fi}
\def\cstok#1{\leavevmode\thinspace\hbox{\vrule\vtop{\vbox{\hrule\kern1pt
\hbox{\vphantom{\tt/}\thinspace{\tt#1}\thinspace}}
\kern1pt\hrule}\vrule}\thinspace}
\centerline {\bf THE IMPACT OF QUANTUM COSMOLOGY ON}
\centerline {\bf QUANTUM FIELD THEORY}
\vskip 1cm
\centerline {GIAMPIERO ESPOSITO}
\vskip 1cm
\centerline {\it Istituto Nazionale di Fisica Nucleare,
Sezione di Napoli}
\centerline {\it Mostra d'Oltremare Padiglione 20,
80125 Napoli, Italy}
\vskip 0.3cm
\centerline {\it Dipartimento di Scienze Fisiche}
\centerline {\it Mostra d'Oltremare Padiglione 19,
80125 Napoli, Italy}
\vskip 1cm
\noindent
{\bf Abstract.} The basic problem of quantum cosmology is
the definition of the quantum state of the universe, with
appropriate boundary conditions on Riemannian three-geometries.
This paper describes recent progress in the corresponding
analysis of quantum amplitudes for Euclidean Maxwell theory
and linearized gravity. Within the framework of
Faddeev-Popov formalism and zeta-function regularization,
various choices of mixed boundary conditions lead to a
deeper understanding of quantized gauge fields and quantum
gravity in the presence of boundaries.
\vskip 100cm
\leftline {\bf 1. Introduction}
\vskip 1cm
\noindent
While the equations of classical field theory are hyperbolic
equations, for which a Cauchy problem can be studied in a
globally hyperbolic space-time, the analysis of Green's
functions of a quantum field theory makes it necessary to
study the so-called Euclidean formulation of field theory.
In the case of Minkowski space-time, the key observation is
that complex Euclidean points belong to the analyticity
domain of the Wightman functions. It is then possible
to define the corresponding Schwinger functions, and a
relativistic quantum field theory can be reconstructed from
a set of Schwinger functions providing the Laplace-transform
condition, Euclidean invariance, reflection positivity,
Euclidean symmetry and cluster property hold [1].

In a curved space-time, however, there is no general result [2]
which ensures that its complexification admits a real section
where the metric is Riemannian, i.e. positive-definite.
Moreover, given a real Riemannian four-manifold, it may be
impossible to recover the Lorentzian Green's functions by
means of analytic continuation. In other words, there is no
smooth way to achieve a transition between a Lorentzian and
a Riemannian formulation of quantum gravity for a generic
space-time. It is instead more appropriate to focus on one
of the two, and then derive all mathematical properties which
are relevant for the given framework. In this paper, we focus
on the elliptic boundary-value problems which are appropriate
for the Riemannian regime. With our terminology, this reduces
to the Euclidean regime in the case of flat background
geometries. The analysis of elliptic operators with suitable
boundary conditions on compact Riemannian three-geometries
makes it possible to give a well-defined formulation of
boundary conditions for quantum gravity and quantum cosmology,
which is one of the main problems of any theory of quantum
gravity, and might shed new light on the problems of
classical cosmology.

Our analysis is concerned with the semiclassical evaluation
of the amplitudes of quantum gravity in the presence of
boundaries [2]. While the path integral for quantum gravity is a
formal object which remains ill-defined, its semiclassical
approximation (although of limited applicability) leads to
well-posed problems with a wide range of applications,
i.e. the one-loop effective action and heat-kernel methods
in field theory, the quantization of constrained systems and
the appropriate boundary conditions for quantum fields.
For a given elliptic operator $\cal A$ (e.g. the Laplace operator
or the squared Dirac operator), its zeta-function
$$
\zeta_{\cal A}(s) \equiv {\rm Tr} \Bigr[{\cal A}^{-s}\Bigr]
\eqno (1.1)
$$
has an analytic continuation to the complex-$s$ plane as a
meromorphic function which is regular at the origin. While
the one-loop effective action depends linearly on
$\zeta_{\cal A}(0)$ and $\zeta_{\cal A}'(0)$ and is non-local,
the scaling properties of the semiclassical amplitudes are
indeed determined by the $\zeta_{\cal A}(0)$ value [2], which
also describes the one-loop divergences of physical theories.

Section 2 describes recent progress on Euclidean Maxwell
theory, and section 3 studies mixed boundary conditions for
Euclidean quantum gravity. Concluding remarks and open
problems are presented in section 4.
\vskip 5cm
\leftline {\bf 2. Euclidean Maxwell Theory}
\vskip 1cm
\noindent
We are interested in the one-loop amplitudes of vacuum
Maxwell theory in the presence of boundaries. Since in the
classical theory the potential $A_{\mu}$ is subject to the
gauge transformations
$$
{\widehat A}_{\mu} \equiv A_{\mu}+\partial_{\mu}\varphi
\; \; \; \; ,
\eqno (2.1)
$$
this gauge freedom is reflected in the quantum theory by a
ghost zero-form, i.e. an anticommuting, complex
scalar field, hereafter denoted again by $\varphi$.
The two sets of mixed boundary conditions
consistent with gauge invariance and BRST symmetry are
magnetic, i.e. [3-4]
$$
\Bigr[A_{k}\Bigr]_{\partial M}=0
\; \; \; \; ,
\eqno (2.2a)
$$
$$
\Bigr[\Phi(A)\Bigr]_{\partial M}=0
\; \; \; \; ,
\eqno (2.2b)
$$
$$
\Bigr[\varphi\Bigr]_{\partial M}=0
\; \; \; \; ,
\eqno (2.2c)
$$
or electric, i.e. [3-4]
$$
\Bigr[A_{0}\Bigr]_{\partial M}=0
\; \; \; \; ,
\eqno (2.3a)
$$
$$
\left[{\partial A_{k}\over \partial \tau}\right]_{\partial M}
=0
\; \; \; \; ,
\eqno (2.3b)
$$
$$
\left[{\partial \varphi \over \partial \tau}\right]_{\partial M}
=0
\; \; \; \; ,
\eqno (2.3c)
$$
where $\Phi$ is an arbitrary gauge-averaging function defined
on the space of connection one-forms $A_{\mu}dx^{\mu}$. Note
that the boundary condition (2.2c) ensures the gauge invariance
of the boundary conditions
(2.2a)-(2.2b) on making the gauge
transformation (2.1). Similarly, the boundary condition (2.3c)
ensures the gauge invariance of (2.3a)-(2.3b) on transforming
the potential as in (2.1).

For a given choice of one of these two sets of mixed boundary
conditions, different choices of background four-geometry,
boundary three-geometry and gauge-averaging function lead to a
number of interesting results. We here summarize them in the
case of a background given by flat Euclidean
four-space bounded by one three-sphere (i.e. the disk) or by
two concentric three-spheres (i.e. the ring).
\vskip 0.3cm
\noindent
(i) The operator matrix acting on the normal and longitudinal
modes of the potential can be diagonalized for all relativistic
gauge conditions which can be cast in the form [3-5]
$$
\Phi_{b}(A) \equiv \nabla^{\mu}A_{\mu}-b \; A_{0} \;
{\rm Tr} \; K
\; \; \; \; ,
\eqno (2.4)
$$
where $\nabla^{\mu}$ denotes covariant differentiation
on the background, $b$ is a dimensionless parameter,
and $K$ is the extrinsic-curvature tensor of the
boundary.
\vskip 0.3cm
\noindent
(ii) In the case of the disk, the Lorentz gauge (set $b=0$
in (2.4)) leads to a $\zeta(0)$ value
$$
\zeta_{L}(0)=-{31\over 90}
\; \; \; \; ,
\eqno (2.5)
$$
for both magnetic and electric boundary conditions, which
agrees with the geometric theory of the asymptotic heat
kernel. However, the $\zeta(0)$ value depends on the gauge
condition, and unless $b$ vanishes it also depends on the
boundary conditions [3-5].
\vskip 0.3cm
\noindent
(iii) In the case of the ring, one finds
$$
\zeta(0)=0
\; \; \; \; ,
\eqno (2.6)
$$
for all gauge conditions [4-6],
independently of boundary conditions
[3-4]. This result agrees with the geometric formulae for the
heat kernel, since volume contributions to $\zeta(0)$ vanish
in a flat background, while surface contributions cancel
each other.
\vskip 0.3cm
\noindent
(iv) In the case of boundary three-geometries given by one
or two three-spheres, the most general gauge-averaging
function takes the form
$$
\Phi(A)=\gamma_{1} \; { }^{(4)}\nabla^{0}A_{0}
+{\gamma_{2}\over 3}A_{0} \; {\rm Tr} \; K
-\gamma_{3} \; { }^{(3)}\nabla^{i}A_{i}
\; \; \; \; ,
\eqno (2.7)
$$
where $\gamma_{1},\gamma_{2}$ and $\gamma_{3}$ are arbitrary
dimensionless parameters. Thus, unless $\gamma_{1},\gamma_{2}$
and $\gamma_{3}$ take some special values (cf. (2.4)), it is
not possible to diagonalize the operator matrix acting on
normal and longitudinal modes of the potential.
\vskip 0.3cm
\noindent
(v) The contributions to $\zeta(0)$ resulting from normal
and longitudinal modes {\it do not} cancel the contribution
of ghost modes, unless one sets $b={2\over 3}$ in (2.4) in
the case of the disk, with magnetic boundary conditions.
Thus, transverse modes do not provide the only contribution
to one-loop amplitudes. In other words, there seem to be
no unphysical modes in a manifestly gauge-invariant quantum
field theory, in that all perturbative modes are necessary to
recover the gauge-invariant quantum amplitudes (see also
section 3).
\vskip 10cm
\leftline {\bf 3. Linearized Gravity}
\vskip 1cm
\noindent
We here focus on the amplitudes of Euclidean quantum gravity
within the framework of Faddeev-Popov formalism. This means
that the amplitudes depend on the boundary data for the
metric and for ghost fields, and are written (formally) as
Feynman path integrals over all compact Riemannian
four-geometries matching the data at the boundary, i.e. [7]
$$
Z[{\rm boundary} \; {\rm data}]
=\int_{C}\mu_{1}[g] \; \mu_{2}[\varphi]
\; e^{-{\widetilde I}_{E}}
\; \; \; \; ,
\eqno (3.1)
$$
where $\mu_{1}$ is a suitable measure on the space of Riemannian
four-metrics, $\mu_{2}$ is a suitable measure on the space of
ghost fields, and the full Euclidean action takes the form
$$ \eqalignno{
{\widetilde I}_{E}&={1\over 16 \pi G}\int_{M}{ }^{(4)}R
\sqrt{{\rm det} \; g} \; d^{4}x
+{1\over 8 \pi G} \int_{\partial M}{\rm Tr} \; K \;
\sqrt{{\rm det} \; q} \; d^{3}x \cr
&+{1\over 32 \pi G \alpha} \int_{M}\Phi_{\nu}\Phi^{\nu}
\; \sqrt{{\rm det} \; g} \; d^{4}x + I_{\rm gh}
\; \; \; \; .
&(3.2)\cr}
$$
With our notation, $q$ is the induced three-metric,
$\Phi_{\nu}$ is a relativistic gauge-averaging function,
and $I_{\rm gh}$ is the corresponding ghost action. In the
one-loop approximation, the measures in (3.1) become measures
on metric perturbations and ghost perturbations, respectively.
Thus, denoting by $g$ the background four-metric and by
$h$ its perturbation, the form of $\Phi_{\nu}$ necessary to
find the familiar form of the propagators, as well as to
recover the Vilkovisky-DeWitt effective action [8-10],
is the de Donder gauge-averaging function [7]
$$
\Phi_{\nu}^{dD}(h) \equiv \nabla^{\mu} \Bigr(h_{\mu \nu}
-{1\over 2}g_{\mu \nu} g^{\rho \sigma}h_{\rho \sigma}\Bigr)
\; \; \; \; ,
\eqno (3.3)
$$
where $\nabla$ is the Levi-Civita connection on the background
four-geometry. The corresponding elliptic operator in the
ghost action is then found to be $-g_{\mu \nu}\cstok{\ }
-R_{\mu \nu}$. Note also that the boundary term in (3.2)
is the one appropriate for fixing the spatial perturbations
$h_{ij}$ at the boundary.

We can now understand how to generalize the magnetic boundary
conditions of section 2 to pure gravity. The basic idea is
to set to zero at the boundary the spatial perturbations
$h_{ij}$ of the metric, and the gauge-averaging function
$\Phi_{\nu}^{dD}(h)$. To ensure that these boundary conditions
are invariant under gauge transformations of $h_{\mu \nu}$ of
the form
$$
{\widehat h}_{\mu \nu} \equiv h_{\mu \nu}
+\nabla_{(\mu} \; \varphi_{\nu)}
\; \; \; \; ,
\eqno (3.4)
$$
one has then to set to zero at the boundary the whole ghost
one-form [11]:
$$
\Bigr[\varphi_{\mu}\Bigr]_{\partial M}=0
\; \; \; \; .
\eqno (3.5)
$$
While $h_{ij}$ obeys homogeneous Dirichlet conditions at
$\partial M$ as we just said, the boundary condition
$\Bigr[\Phi_{\nu}^{dD}(h)\Bigr]_{\partial M}=0$ leads to [11]
$$
\left[{\partial h_{00}\over \partial \tau}
+{6\over \tau}h_{00}-{\partial \over \partial \tau}
\Bigr(g^{ij}h_{ij}\Bigr)+{2\over \tau^{2}}
h_{0i}^{\; \; \; \mid i} \right]_{\partial M}=0
\; \; \; \; ,
\eqno (3.6)
$$
$$
\left[{\partial h_{0i}\over \partial \tau}
+{3\over \tau}h_{0i}-{1\over 2}{\partial h_{00}\over
\partial x^{i}}\right]_{\partial M}=0
\; \; \; \; .
\eqno (3.7)
$$
In the case of flat Euclidean four-space bounded
by a three-sphere, which is relevant for quantum
cosmology in the case of four-sphere backgrounds bounded
by a three-sphere of small radius [2],
the boundary conditions described so far,
which were first proposed by Barvinsky [12], lead to the
full $\zeta(0)$ value [11]
$$
\zeta(0)=-{241\over 90}
\; \; \; \; .
\eqno (3.8)
$$
This differs from the contribution of transverse-traceless
metric perturbations, which was found to be [13]
$$
\zeta_{{\rm TT}}(0)=-{278\over 45}
\; \; \; \; .
\eqno (3.9)
$$
Interestingly, the detailed calculations performed in [11]
and outlined in this section add evidence in favour of no
cancellation being possible between ghost- and gauge-modes
contributions to one-loop amplitudes in the presence of
boundaries. From the point of view of constrained Hamiltonian
systems and their quantization, this seems to suggest that
there are no unphysical modes in a gauge-invariant quantum
field theory. Reduction of a field theory with first-class
constraints to its physical degrees of freedom before
quantization leads to an inequivalent quantum field theory,
where gauge-invariance properties are lost (cf. [14]).

The boundary conditions studied in [11-12] are not the only
possible set of mixed boundary conditions for Euclidean
quantum gravity [15].
By contrast, on studying BRST transformations
at the boundary, one is led to
consider the following boundary conditions [11,16-17]:
$$
\left[\Bigr(2{\rm Tr} \; K + n^{\alpha}\nabla_{\alpha}\Bigr)
n^{\mu}n^{\nu}\Bigr(h_{\mu \nu}-{1\over 2}
g_{\mu \nu}g^{\rho \sigma}h_{\rho \sigma}\Bigr)
\right]_{\partial M}=0
\; \; \; \; ,
\eqno (3.10)
$$
$$
\Bigr[h_{ij}\Bigr]_{\partial M}
=\Bigr[h_{0i}\Bigr]_{\partial M}
=\Bigr[\varphi_{0}\Bigr]_{\partial M}=0
\; \; \; \; ,
\eqno (3.11)
$$
$$
\left[\Bigr(-K_{\mu}^{\; \; \nu}
+\delta_{\mu}^{\; \; \nu} \; n^{\alpha}\nabla_{\alpha}
\Bigr)P_{\nu}^{\; \; \sigma} \; \varphi_{\sigma}
\right]_{\partial M}=0
\; \; \; \; ,
\eqno (3.12)
$$
where there is summation over repeated indices, and we
have used the tangential projection operator
$$
P_{\mu}^{\; \; \nu} \equiv \delta_{\mu}^{\; \; \nu}
-n_{\mu} \; n^{\nu}
\; \; \; \; ,
\eqno (3.13)
$$
$n^{\mu}$ being the normal to the boundary.

As shown in [11], the boundary conditions (3.10)-(3.12)
lead to the following $\zeta(0)$ value in the case of
flat Euclidean space bounded by $S^{3}$:
$$
\zeta(0)=-{758\over 45}
\; \; \; \; ,
\eqno (3.14)
$$
which {\it agrees} with the results deriving from the
geometric theory of the asymptotic heat kernel [11,18-20].
By contrast, the boundary conditions (3.5)-(3.7) make it
more difficult to use projection operators and then apply
the powerful geometric techniques available in the literature.
Nevertheless, the St. Petersbourg group, led by Dr. D.
Vassilevich, is making progress on this crucial issue.
\vskip 1cm
\leftline {\bf 4. Open Problems}
\vskip 1cm
\noindent
The analysis of Euclidean Maxwell theory in the presence
of boundaries raises at least three crucial issues. First,
since in the one-boundary problems the Faddeev-Popov amplitudes
turn out to be gauge-dependent, should we accept that not
all gauges are admissible, or should we instead argue that the
Hartle-Hawking program [21] is incorrect, because one cannot shrink
to a point one of the two boundary three-surfaces ? Second,
how to prove explicitly the gauge invariance of quantum amplitudes
in the two-boundary problems. What happens is that changing the
gauge condition leads to a smooth variation of the matrix of
elliptic operators acting on gauge modes [7]. One has then to prove
that the resulting contributions to $\zeta(0)$ remain unaffected
by such a smooth variation, even though it is no longer possible
to express the gauge modes as linear combinations of Bessel
functions. Third, the lack of cancellation of gauge- and
ghost-modes contributions to the full $\zeta(0)$ points out a
deeper role played by such modes in the quantum theory. They
are essential to recover the full content of the semiclassical
approximation, and hence cannot be regarded as non-physical,
although in the classical Lorentzian theory one is naturally led
to identify the transverse part of the electromagnetic
potential with the physical degrees of freedom [14].

In the case of linearized Euclidean quantum gravity, the explicit
proof of gauge invariance of the one-loop amplitudes is even
more complicated, since there are now ten sets of perturbative
modes. Moreover, it appears necessary to obtain geometric
formulae for the asymptotic heat kernel in the case of
Barvinsky boundary conditions [12] studied in section 3. Other
relevant problems are the analysis of non-relativistic gauges
for pure gravity, and the non-local nature of the one-loop
effective action expressed through the
$\zeta'(0)$ value for elliptic problems with boundaries [22].

Last, but not least, the quantum state of the Lorentzian
theory corresponding to the boundary conditions of section 3
remains unknown. If this problem is not thoroughly studied,
we remain unable to make contact with the world we live in,
unless one is ready to accept Hawking's view, according to
which the Euclidean regime is the more fundamental [23].

In the light of the analysis presented in this paper, it
seems appropriate to conclude that quantum cosmology has
indeed a deep influence on the understanding of physical
fields and their quantization, and hence it lies at the
very heart of fundamental theoretical physics.
\vskip 1cm
\leftline {\bf Acknowledgments}
\vskip 1cm
\noindent
I am much indebted to Alexander Kamenshchik, Igor Mishakov
and Giuseppe Pollifrone for collaboration on the topics
described in this paper, and to Andrei Barvinsky and Dmitri
Vassilevich for correspondence on mixed boundary conditions
for Euclidean quantum gravity. This research received financial
support by the Istituto Nazionale di Fisica Nucleare, and
by the European Union under the Human Capital and
Mobility Program.
\vskip 1cm
\leftline {\bf References}
\vskip 1cm
\item {[1]}
Strocchi F. (1993) {\it Selected Topics on the General Properties
of Quantum Field Theory} (Singapore: World Scientific).
\item {[2]}
Esposito G. (1994) {\it Quantum Gravity, Quantum Cosmology
and Lorentzian Geometries}, Lecture Notes in Physics,
New Series m: Monographs, Vol. m12, second corrected and
enlarged edition (Berlin: Springer-Verlag).
\item {[3]}
Esposito G. (1994) {\it Class. Quantum Grav.} {\bf 11}, 905.
\item {[4]}
Esposito G., Kamenshchik A. Yu., Mishakov I. V. and
Pollifrone G. (1994) {\it Class. Quantum Grav.} {\bf 11}, 2939.
\item {[5]}
Esposito G., Kamenshchik A. Yu., Mishakov I. V. and
Pollifrone G. {\it Relativistic Gauge Conditions in
Quantum Cosmology} (DSF preprint 95/8, to
appear in {\it Phys. Rev.} {\bf D}).
\item {[6]}
Esposito G. and Kamenshchik A. Yu. (1994) {\it Phys. Lett.}
{\bf B 336}, 324.
\item {[7]}
Esposito G., Kamenshchik A. Yu., Mishakov I. V. and
Pollifrone G. (1994) {\it Phys. Rev.} {\bf D 50}, 6329.
\item {[8]}
Taylor T. R. and Veneziano G. (1990) {\it Nucl. Phys.}
{\bf B 345}, 210.
\item {[9]}
Vilkovisky G. A. (1984) {\it Nucl. Phys.} {\bf B 234}, 125.
\item {[10]}
De Witt B. S. (1987) {\it The Effective Action}, in
{\it Architecture of Fundamental Interactions at Short
Distances}, Les Houches Session XLIV, eds. P. Ramond and
R. Stora (Amsterdam: North-Holland) p. 1023.
\item {[11]}
Esposito G., Kamenshchik A. Yu., Mishakov I. V. and
Pollifrone G. {\it One-Loop Amplitudes in Euclidean
Quantum Gravity} (DSF preprint 95/16).
\item {[12]}
Barvinsky A. O. (1987) {\it Phys. Lett.} {\bf B 195}, 344.
\item {[13]}
Schleich K. (1985) {\it Phys. Rev.} {\bf D 32}, 1889.
\item {[14]}
McMullan D. and Tsutsui I. (1995) {\it Ann. Phys.}
{\bf 237}, 269.
\item {[15]}
Esposito G. and Kamenshchik A. Yu. {\it Mixed Boundary
Conditions in Euclidean Quantum Gravity} (DSF preprint
95/23).
\item {[16]}
Luckock H. C. (1991) {\it J. Math. Phys.} {\bf 32}, 1755.
\item {[17]}
Moss I. G. and Poletti S. J. (1990) {\it Nucl. Phys.}
{\bf B 341}, 155.
\item {[18]}
Branson T. P. and Gilkey P. B. (1990) {\it Commun. Part.
Diff. Eq.} {\bf 15}, 245.
\item {[19]}
Vassilevich D. (1995) {\it J. Math. Phys.} {\bf 36}, 3174.
\item {[20]}
Moss I. G. and Poletti S. J. (1994) {\it Phys. Lett.}
{\bf B 333}, 326.
\item {[21]}
Hartle J. B. and Hawking S. W. (1983) {\it Phys. Rev.}
{\bf D 28}, 2960.
\item {[22]}
Bordag M., Geyer B., Kirsten K. and Elizalde E.
{\it Zeta-Function Determinant of the Laplace Operator
on the D-Dimensional Ball} (UB-ECM-PF preprint 95/10).
\item {[23]}
Gibbons G. W. and Hawking S. W. (1993) {\it Euclidean
Quantum Gravity} (Singapore: World Scientific).

\bye